# Large-Scale Dynamos Driven by Shear-Flow-Induced Jets

B. Tripathi[1,*], A. E. Fraser[2], P. W. Terry[1], E. G. Zweibel[1], M. J. Pueschel[3,4,5] & R. Fan[1]

[1]University of Wisconsin–Madison, Madison, Wisconsin 53706, USA.
[2]University of Colorado, Boulder, Colorado 80309, USA.
[3]Dutch Institute for Fundamental Energy Research, 5612 AJ Eindhoven, The Netherlands.
[4]Eindhoven University of Technology, 5600 MB Eindhoven, The Netherlands.
[5]Department of Physics & Astronomy, Ruhr-Universität Bochum, 44780 Bochum, Germany.
[*]btripathi@wisc.edu

**At every scale they occupy, magnetic fields affect various phenomena, including star formation, cosmic ray transport, charged particle acceleration, space weather, transport in planetary atmospheres, and laboratory plasmas. These fields are often generated and sustained by turbulent flows in a process called the dynamo. In 1955, E. N. Parker parameterized the effects of small-scale turbulence to propose a mean-field dynamo theory[1]. The widely used theory reproduces observed large-scale fields but suffers from difficulty in tuning parameters as they are not justified from first principles: Studies of turbulent flows show tangled magnetic fields, which are folded and fragmented into small-scale structures due to shear-flow straining[2,3]. Here, considering a shear flow that is unstable and driven, we develop analytic theory and perform three-dimensional (3D), advanced computer simulations of turbulence with up to 4096 × 4096 × 8192 grid points, showing *ab initio* generation of quasi-periodic, large-scale magnetic fields. The generation occurs via the mean-vorticity effect—an additional mean-field dynamo process postulated[4] in 1990. Crucial to this dynamo is the prior generation of large-scale 3D jets, robustly produced as topologically protected and exact nonlinear solutions of the magnetohydrodynamic equations. The jet-driven dynamo applies to shear-driven laboratory and astrophysical systems. These include binary neutron star mergers[5,6], where the reported dynamo likely operates on microsecond timescales to produce in milliseconds some of the strongest magnetic fields in the Universe[7], providing signals for multi-messenger astronomy[8].**

Flows with gradients are very common in nature; they are often unstable, drive turbulence, and appear with magnetic fields. Examples are found in laboratory experiments, planetary atmospheres and interiors, the meridional circulation and near-surface shear layers of the Sun, stellar interiors, accretion disks and jets, binary neutron star mergers, galaxy clusters, and galactic rotation curves[5,9–18]. Shear flows abet small-scale structures[2,3], especially when the flow is unstable, as the cascade of the flow-energy originating at large scale generates disordered small-scale turbulent fluctuations. A long-standing challenge has been to explain how the *large-scale*, smooth magnetic fields—observed in magnetized planets, the solar cycle, stars, galaxies, and cosmic voids—can arise in the presence of destructive turbulent motions[9,19–23].

Traditional dynamos constrained by magnetic helicity generally fail to generate large-scale fields unless they meet certain strict requirements[1,20,24–27]. Even then, those dynamos have not proved robust because of effects like Alfvénization—a fundamental magnetohydrodynamic (MHD) process via which velocity and magnetic fluctuations becomes equipartitioned and aligned. Because of this alignment, the traditional dynamos are almost completely suppressed[28]. To help explain observations, we investigate here a fundamentally different dynamo mechanism where Alfvénization *enables* the generation of large-scale magnetic fields. This postulated mechanism linked to large-scale vortical structures[4] has remained hypothetical. We find that this hypothetical mechanism develops from the self-organization of large-scale jets. We answer here whether this dynamo can arise naturally in astrophysical environments[5,9–16], by what mechanisms it operates, and how robust it is. Because large-scale self-organized jets are often present in vortex-dominated systems[29], we also address the relationship of the former to the latter, and how they work together to generate large-scale fields.

We analytically and numerically demonstrate a robust non-traditional dynamo effect. The dynamo ensues regardless of whether the domain considered is non-periodic or periodic, and whether an initial large-scale external field is present or absent. We consider a 3D domain with a large-scale flow

$$\mathbf{u}_0=U_x(z)\hat{\mathbf{e}}_x, \qquad (1)$$

where $\hat{\mathbf{e}}_x$ represents the unit vector along $x$, and $U_x(z)$ is a $z$-varying profile. We define the $(x,y)$-averaged profile as the mean. When the mean shear flow is unstable [Kelvin–Helmholtz (KH) instability[10]], it drives turbulence across a range of scales, including small scales. The small scales saturate over some multiple of the instability growth time $\tau_{\text{grow}}\approx 5\ a/U_0$, for $U_x(z)=U_0 \tanh(z/a)$, where $a$ is the half-width of the shear flow with amplitude $U_0$. The evolution of a large-scale dynamo, however, requires longer times and hence demands substantial computational effort. On the other hand, since turbulence depletes the large-scale flow gradient over time, dynamo eventually becomes inactive. Hence, we maintain the large-scale flow $U_x(z)$ externally[30], mimicking persistent astrophysical shear flows (Methods Sec. V); and we simulate three-dimensional (3D) incompressible MHD turbulence by extending most simulations to ≈350 $\tau_{\text{grow}}$ and one simulation to 2,400 $\tau_{\text{grow}}$ (see Methods Sec. I).

Figure 1 displays the temporal evolution of the $(x,y)$-averaged mean field, turbulent fields, and turbulent flows. The initial horizontal field $\mathbf{b}_0$ in this simulation is weak and uniform throughout the domain [$U_0/|\mathbf{b}_0|=30$; $\measuredangle(\mathbf{u}_0, \mathbf{b}_0)=30^\circ$]. The mean field generated by the turbulence, however, is drastically different: the mean-field energy is amplified by 3 orders of magnitude; the mean field is almost entirely (anti-)aligned to the mean flow; and the mean field is reversed across $z$. The reversed mean field changes its polarity quasi-periodically in Fig. 1**a**, which is reminiscent of the solar magnetic cycles[31]. The generation of the $x$-directed, $z$-reversed mean field is unexpected, because here the mean magnetic field cannot be directly generated by the mean flow. This is revealed in the electromotive force (EMF, $\boldsymbol{\mathcal{E}}$)

$$\langle\boldsymbol{\mathcal{E}}\rangle_{x,y} = \underbrace{\langle\mathbf{U}\rangle_{x,y}\times\langle\mathbf{B}\rangle_{x,y}}_{\text{zero }\Omega-\text{effect}} + \underbrace{\langle\tilde{\mathbf{u}}\times\tilde{\mathbf{b}}\rangle_{x,y}}_{\text{dominant }\Upsilon-\text{effect}}, \qquad (2)$$

where the angular brackets $\langle\cdot\rangle_{x,y}$ average fluctuations in the $(x,y)$-plane; $\mathbf{U}$ and $\mathbf{B}$ respectively represent the $(x,y)$-averaged mean flow and mean magnetic fields; $\tilde{\mathbf{u}}$ and $\tilde{\mathbf{b}}$ respectively represent turbulent flow and turbulent magnetic fields; thus, $\mathbf{U}+\tilde{\mathbf{u}}$ and $\mathbf{B}+\tilde{\mathbf{b}}$ represent the total velocity and total magnetic field. In the curl of Eq. (2), the second term produces straining of mean field by mean flow (the so-called dynamo Ω-effect[32]), which is zero here. That is, $\mathbf{B}\cdot\nabla U_x=(B_x\partial_x+B_y\partial_y)U_x=0$ [for the $(x,y)$-averaged mean flow and fields, $\nabla\cdot\mathbf{B}=\partial_z B_z=0$ and $\partial_z U_z=0$; the mean vertical field $B_z$ and flow $U_z$ are absent in the present simulations]. Thus, the $(x,y)$-averaged mean field *must* be generated by turbulent effects via the third term of Eq. (2).

The dominant turbulent flows are jet-like because they are directed along $x$ and do not vary along $x$ (similar to axisymmetric azimuthal flows in planetary atmospheres); see the red-blue 3D streamlines of $\tilde{u}_x$ in Figs. 1**c**,**d**. This jet-like flow $\tilde{u}_x$ exhibits even-parity symmetry along $z$. Since $\partial_z\tilde{u}_x$ changes sign along $z$, the field-line-stretching $\tilde{\mathbf{b}}\cdot\nabla\tilde{u}_x$ by turbulent jets generates a $z$-reversed mean field. This is further explained by a semi-local analytic theory (Methods Sec. VIII). The origin of the jets is non-trivial, as perturbations with the wavenumber $k_x=0$ are linearly stable (Extended data Fig. 1). The jets are formed when the mean flow horizontally stretches a seed fluctuation flow $\tilde{u}_z$ (Methods Sec. VII). The seed fluctuation flow is excited by the KH instability nonlinearly. The jets and seed fluctuation flow have magnetic counterparts.

The large-scale jets $\tilde{u}_x$ are robust to variations in fluid Reynolds number $Re$ ($\propto 1/\text{viscosity}$) and magnetic Reynolds number $Rm$ ($\propto 1/\text{resistivity}$); see Fig. 2**a**. These $x$-directed, $x$-invariant jets are exact solutions to the *nonlinear* ideal MHD equations—akin to the Elsässer fields[33] and zonal flows[13]. These structures render all MHD nonlinearities zero. The spatial topology of the jets protects them from being destroyed because the gradients of perturbed fluid pressure and magnetic pressure are zero.

The jets operate on the seed fluctuation magnetic field $\tilde{b}_z$ to induce a dynamo (Methods Sec. VIII). Both of these vary in the $(y,z)$-plane, but not in $x$. (In Fig. 2**a**, $\tilde{b}_z$ is evaluated for $k_x=0$, $k_y=2\pi/L_y$, with $L_y=10\pi$.) The amplitude (square root of energy) of the dynamo-generated mean field $B_x$ has weak sensitivity to variations in $Pm$—the ratio of viscosity to resistivity; see Fig. 2**b**. This behavior is due to the robust jets $\tilde{u}_x$, as the jets operate on $\tilde{b}_z$ via $\langle\tilde{\mathbf{b}}\cdot\nabla\tilde{u}_x\rangle_{x,y}=\partial_z\langle\tilde{b}_z\tilde{u}_x\rangle_{x,y}$ to generate the mean field $B_x$.

Traditional mean-field dynamo theory expresses the turbulent EMF $\boldsymbol{\mathcal{E}}$ in terms of the mean magnetic field $\mathbf{B}$, by removing a uniform mean flow via Galilean transformation[24]. In the presence of an inhomogeneous mean flow $\mathbf{U}$, which cannot be removed, traditional mean-field theory needs to be generalized[4]. The essential elements of this generalization are reproduced by our quasilinear EMF model (Methods Sec. VI). This generalized theory predicts $\boldsymbol{\mathcal{E}}=\alpha\mathbf{B}-\beta\nabla\times\mathbf{B}+\Upsilon\nabla\times\mathbf{U}$, where $\alpha$ captures kinetic helicity $\langle\tilde{\mathbf{u}}\cdot\nabla\times\tilde{\mathbf{u}}\rangle_{x,y}$ and current helicity $\langle\tilde{\mathbf{b}}\cdot\nabla\times\tilde{\mathbf{b}}\rangle_{x,y}$; here, $\beta$ measures turbulent energy; and the Upsilon $\Upsilon$ stands for the Yoshizawa's postulated mean-vorticity effect[4] related to the cross-helicity $\langle\tilde{\mathbf{u}}\cdot\tilde{\mathbf{b}}\rangle_{x,y}$. The $\Upsilon$-effect is robust to variations in $Pm$ (Fig. 2**c**; Methods Sec. VI). Similar trend of $\Upsilon$ shown during field-growth is found in the saturated phase. In addition to the multidimensional spatial regression[27] used in Fig. 2**c**, we have employed temporal regression[34] to recover the dominance of the $\Upsilon$-effect.

We develop a physical understanding of the $\Upsilon$-effect in Fig. 3**a**. This effect has a strong analogy with the $\alpha$-effect[1]. A visual representation of the $\alpha$-effect ($\alpha\mathbf{B}\propto\mathbf{B}\langle\tilde{\mathbf{u}}\cdot\nabla\times\tilde{\mathbf{u}}\rangle$) involves three elements: magnetic field ($\mathbf{B}$), velocity ($\tilde{\mathbf{u}}$), and vorticity ($\nabla\times\tilde{\mathbf{u}}$). Following that order, an initial straight mean field $\mathbf{B}$ is bent and rotated[1], thus creating a mean field orthogonal to $\mathbf{B}$. To similarly visualize the ϒ-effect ($\Upsilon\nabla\times\mathbf{U}\propto\nabla\times\mathbf{U}\langle\tilde{\mathbf{u}}\cdot\tilde{\mathbf{b}}\rangle$), we reverse the order of the previous three elements: vorticity ($\nabla\times\mathbf{U}$), velocity ($\tilde{\mathbf{u}}$), and magnetic field ($\tilde{\mathbf{b}}$). Following thar order, an initial straight mean vortex line is bent, inducing jets (Fig. 3**a**). The jets are 3D as they vary along the *y*-direction. The jets then operate on $\tilde{\mathbf{b}}$ to produce a mean field.

Our proposed dynamo mechanism is confirmed in Fig. 3**b** using a computer simulation of an $(x,y,z)$-domain with $4096\times4096\times8192$ grid points. The equilibrium flow here is a double shear layer[35] (Methods Secs. II, III). This calculation is, to our knowledge, the highest-resolution Kelvin–Helmholtz spectral dynamo simulation to date (Methods Sec. IV). In the beginning of this simulation, only the mean flow is present—and *the mean magnetic field is zero*. We add perturbations in velocity and magnetic field with energy $10^{-18}$ times the mean-flow energy. The physical mechanism presented in Fig. 3 dominantly generates mean magnetic field in all simulations where the mean flow is maintained (Methods Sec. XII).

Figure 4**a** shows vorticity and electric current density, which are aligned because velocity and magnetic fields are aligned, as measured in Fig. 4**b**. Unlike energies, helicities can have different signs at different scales[36]. The dynamo features a cascade of energy from large scales to small scales (Extended data Fig. 2; Methods Sec. IX). Small scales take energy away from the mean field; the large-scale jets give energy to the mean field (Extended data Fig. 3).

The first direct laboratory measurement of a turbulent EMF in the Madison Dynamo Experiment (MDE) presented a finding[37] that challenged traditional theories: The magnetic field and the EMF in the experiment were nearly orthogonal (Extended data Fig. 4), showing that the $\alpha$-effect fails to explain their dynamo. Their measurement is consistent with the ϒ-effect if one considers the observed large-scale radial vorticity[38], which drives the observed radial EMF. In both MDE and our simulations, large-scale vortical motions are replenished by external forces. Both systems host turbulently generated seed fluctuation flows and jets. Both systems exhibit the same alignment of different components of EMF. This work suggests that the large-scale jet-driven dynamos are present in the laboratory and nature[39] (Methods Sec. X).

The classical relaxed states of MHD turbulence are force-free[40]; these states in $\alpha$-dynamos have $\alpha\mathbf{B} - \beta\nabla\times\mathbf{B} = 0$. However, KH-dynamo simulations show that $\mathbf{B}$ and $\nabla\times\mathbf{B}$ are orthogonal, as the mean magnetic field $\mathbf{B}$ is reversed along $z$ and directed dominantly along $x$. The relaxed KH-dynamo states feature $\mathbf{B}$ (anti-)parallel to $\mathbf{U}$ because the inductive term takes the form $\nabla\times(\Upsilon\nabla\times\mathbf{U})$. The sign of ϒ determines the polarity of $\mathbf{B}$.

Binary neutron stars (BNS), as they approach each other, produce in the merger interface a KH-unstable velocity gradient[5,7], where a non-relativistic treatment captures the essential physics. In the KH-unstable system, the field grows due to the ϒ-dynamo—as seen in simulations—at a typical rate of $0.04\ U_0/a$, where $a$ is the half-width of the shear flow with an amplitude $U_0$. Measurements from general-relativistic MHD simulations of BNS mergers show that the flows in the thin merger interface[5,6] have $a$~10–15 m (likely even smaller[11]) and $U_0$~$0.1c$, with $c$ representing the speed of light. Hence, we predict the e-folding time of the ϒ-dynamo in BNS mergers to be ~8 μs (variations depend on the model, but this timescale is likely even shorter).

The BNS merger takes milliseconds[5], during which the shear layer persists between the approaching BNS. Thus, the ϒ-dynamo can create large-scale magnetic fields even from zero initial large-scale field and rapidly amplify them to energies similar to the bulk kinetic energy[7]. These fields, possibly[41] of $10^{16}$–$10^{17}$ G, impact the merger and post-merger dynamics, and change the gravitational waveforms by increasing their frequencies[8] by 200–300 Hz. These changes are in principle detectable by the LIGO-Virgo-KAGRA Collaboration in upcoming runs[42] and will be measured by the Einstein telescope[43]. Strong magnetic fields affect electron acceleration and synchrotron radiation, thereby impacting the electromagnetic signals[41]. High-resolution space observations and global merger simulations[5,6] can capture the rapid generation of magnetic fields via the ϒ-effect (Methods Sec. XI). Sub-grid models[44] should be able to reproduce the ϒ-dynamo driven by large-scale jets. The ϒ-dynamo also has implications for merging galaxy clusters[17], where large-scale unstable shear flows exist and large-scale magnetic fields have been reported[9,18].

The jet-driven vorticity-related dynamo has bearing for the generation of magnetic fields in the Sun. The solar meridional flow—a key player in flux transport dynamos[14]—has vorticity directed in the azimuthal direction[15]. This vorticity is bound to drive the azimuthal component of the turbulent electromotive force, generating poloidal magnetic fields. The large vorticity of solar subsurface flows[45] is also expected to generate magnetic fields. Simulations show large cross-helicity in the near-surface shear layers, which have poloidal vorticity[46]. Relatedly, zonal jets and torsional oscillations in the outer parts of the solar interior are precursors of the solar magnetic cycles[29,45,47]. Simpler models of the ϒ-dynamo are useful to explain stellar magnetic cycles[48]. The jet-driven dynamo is expected to generate magnetic fields in planets and galaxies, as well, which host vortical structures[16]. Since common MHD turbulence theories apply only in the case of zero global cross-helicity, this work motivates the understanding of MHD turbulence theories for non-zero global cross-helicity[49] (Extended data Fig. 5).

This article has outlined the salient features of the jet-driven ϒ-dynamo arising in turbulence generated by shear flows, with potential applications in laboratory dynamo experiments, the Sun, main-sequence stars, planets, accretion flows around compact objects, and binary neutron star mergers. The robust nonlinear jets are shown here to be the physical mechanism behind the previously postulated ϒ-effect[4], which this work confirms.

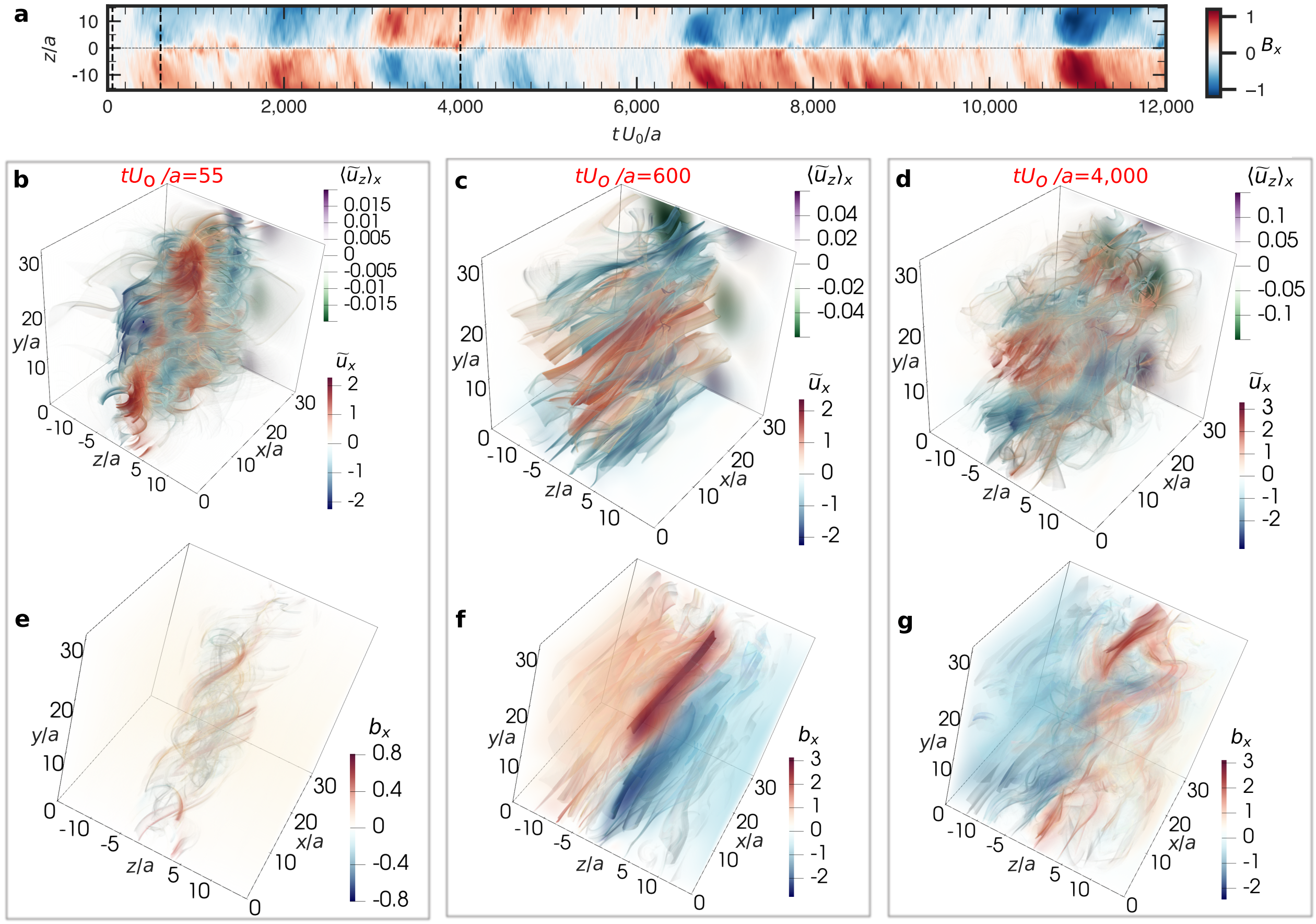

**Fig. 1 | Long-time evolution of the MHD-KH dynamo, displaying phases of turbulent generation of flows and magnetic fields. a**, The $(x,y)$-averaged mean field $B_x$ quasi-cyclically reverses ($t \approx 3{,}000$; 5,300; 6,000; 6,350). **b**–**d**, 3D streamlines in red-blue show turbulent velocity [the mean flow $U_x(z)$ is not shown]. Colors correspond to $\tilde{u}_x$, at times indicated by vertical dashed lines in **a**. The back-plane at $x$=10π displays vertical velocity $\langle\tilde{u}_z\rangle_x$, which is Fourier-filtered to obtain a fluctuation that is invariant in $x$ and fundamental harmonic in $y$. At $t$=55 in **b**, the two-dimensional KH eddies appear as cylinders, with almost no variation along $y$, but they have transformed into $x$-invariant jets in **c** and **d**. These jets are anchored on the $(y,z)$-plane at $x$=10π, where $\langle\tilde{u}_z\rangle_x$ is large, shown with a purple-green color bar. **e**–**g**, Streamlines of magnetic fields are overlaid on top of volume-rendered volume-filling fields. The polarities of the mean fields flip between **f** and **g**.

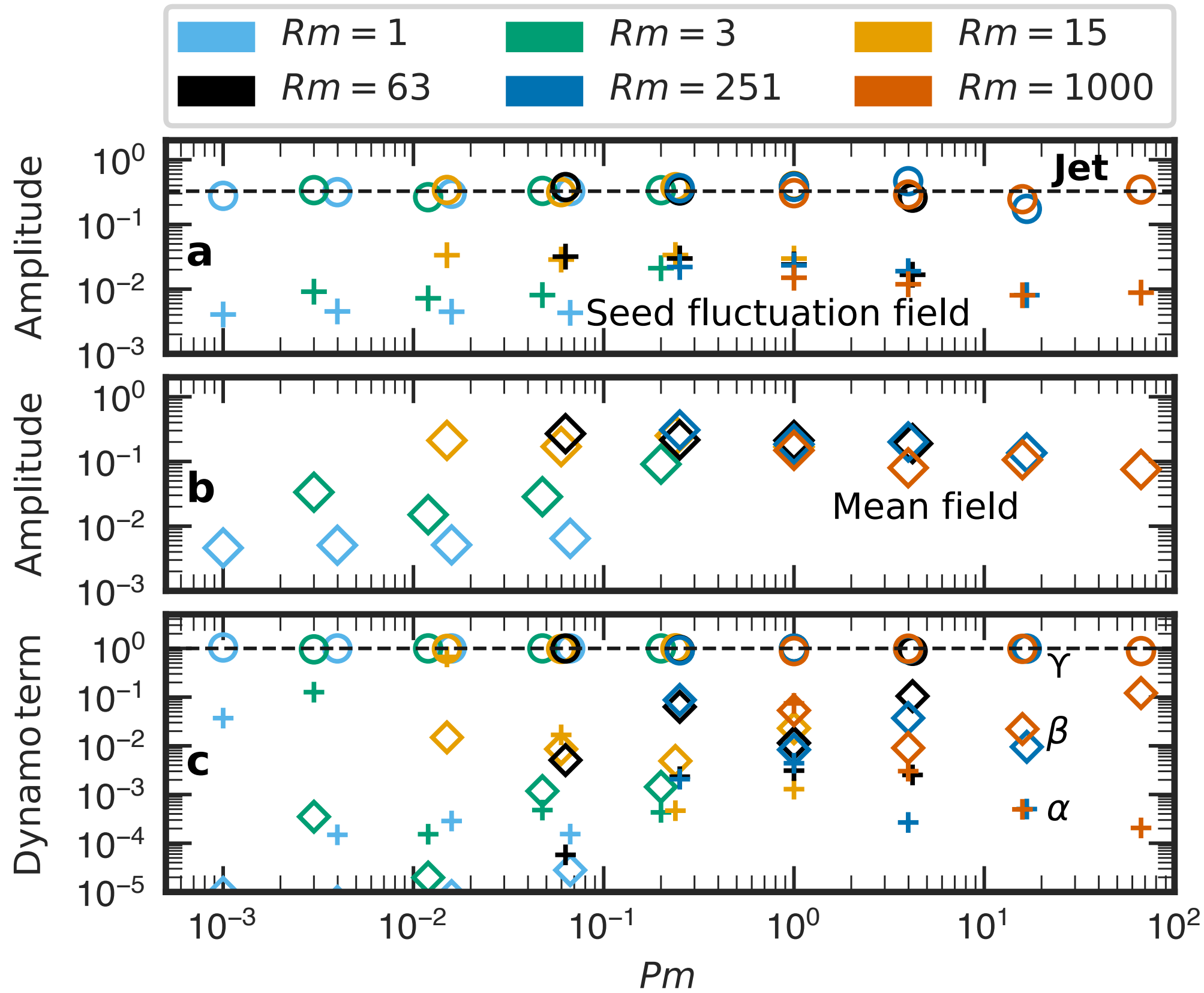

**Fig. 2 | Properties of the KH instability-induced jets and dynamo. a**, Amplitudes of the 3D jets (shown with circles) remain unchanged over variations in magnetic Prandtl number *Pm*. Reynolds numbers *Re* and *Rm* are measured using length *a* and speed $U_0$ of the mean flow. Above the threshold $Rm_c$ between 3 and 15, the seed fluctuation field (plus sign) induced by the KH instability does not have drastic variation. For every *Rm*, we show four cases: *Re*=15, 63, 251, and 1000, which are indicated by four circles and four pluses of same color. Thus, amplitudes are observed to be independent of *Re*. **b**, The dynamo-generated mean field $B_x$ does not have strong sensitivity to visco-resistive properties above $Rm_c$. This is a consequence of the results shown in panel **a**. **c**, The mean turbulent EMF is driven by the Υ-effect. Circles, diamonds, and pluses represent the time-averaged contributions of the Υ-, *β*-, and *α*-effects to the mean EMF, quantified by $|\Upsilon(\nabla\times\mathbf{U})_y/\mathcal{E}_y|$, $|\beta(\nabla\times\mathbf{B})_y\,/\mathcal{E}_y|$, and $|\alpha B_y/\mathcal{E}_y|$, respectively. Similar dominant Υ-effect is found using different time windows. The EMF is the largest where the mean flow reverses.

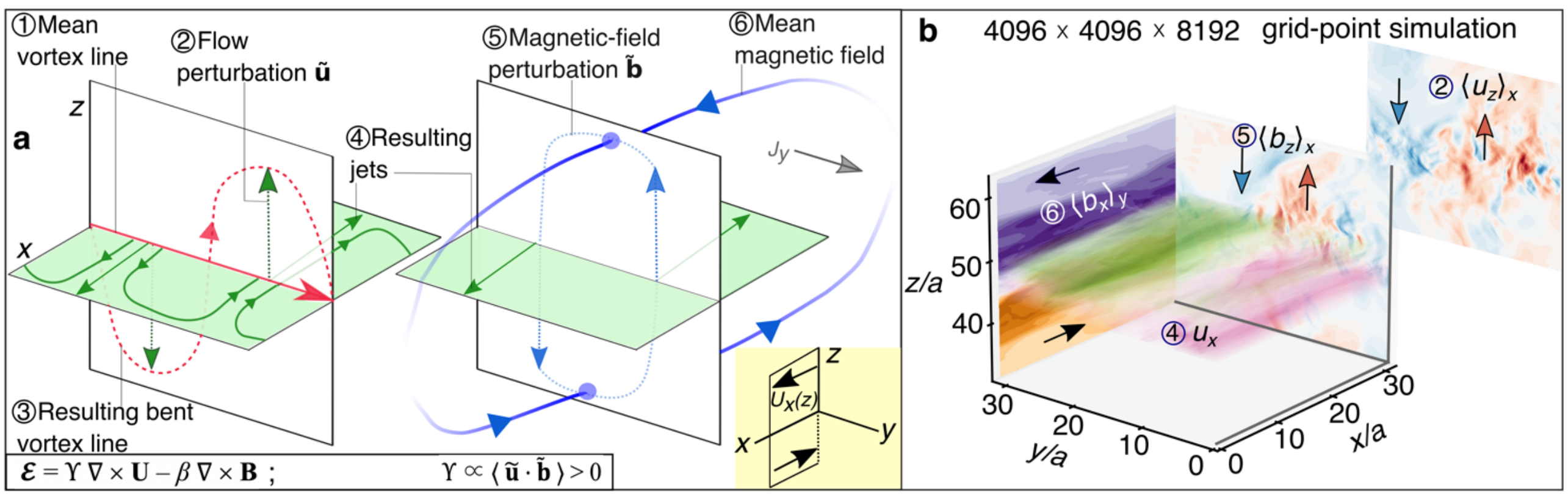

**Fig. 3 | Mechanism of the *jet-driven* ϒ-dynamo. a**, Red, green, and blue represent vorticity, velocity, and magnetic field, respectively. Step ① shows a *y*-directed mean vortex-line, corresponding to a mean shear flow $U_x(z)$ in the $(x,z)$-plane, shown at the bottom right of panel **a**. In step ②, the KH instability excites a flow perturbation $\tilde{\mathbf{u}}$, with a sinusoidal variation in *y*, representing a purely 3D process. This flow perturbation bends the mean vortex line in step ③. Where the vortex line points upward, eddy-like flows are induced around it; these are shown with solid green curves in step ④. This hydrodynamic process generates jets, which have no variation along *x* but are directed along *x*. Step ⑤ is the magnetic counterpart of step ②; the alignment between $\tilde{\mathbf{u}}$ and $\tilde{\mathbf{b}}$ represents here the case of positive ϒ (an Alfvénic state). The *y*-varying jets of step ④ stretch the *y*-varying magnetic-field perturbation of step ⑤, thereby creating a mean magnetic field (invariant in *x* and *y*) in step ⑥. The mean field aligns with the mean flow. **b,** All quantities are plotted using red (pink) for positive and blue (green) for negative, where the color map spans [-*M*,*M*], with *M* representing the absolute maximum of each quantity. A half of the symmetric double shear layer[35] is shown. Since flow and fields in steps ② and ⑤ (in panel **a)** do not vary along *x*, they are *x*-averaged and displayed on the $(y,z)$-plane at arbitrary *x*-coordinates for visual clarity. Multiple $(x,y)$-slices of jets $u_x$ are plotted near a shear layer interface where jets are the strongest. The *x*-directed magnetic field is averaged in *y*, and $\langle b_x \rangle_y$ shown on the $(x,z)$-plane. The mean flow is shown with the black arrows.

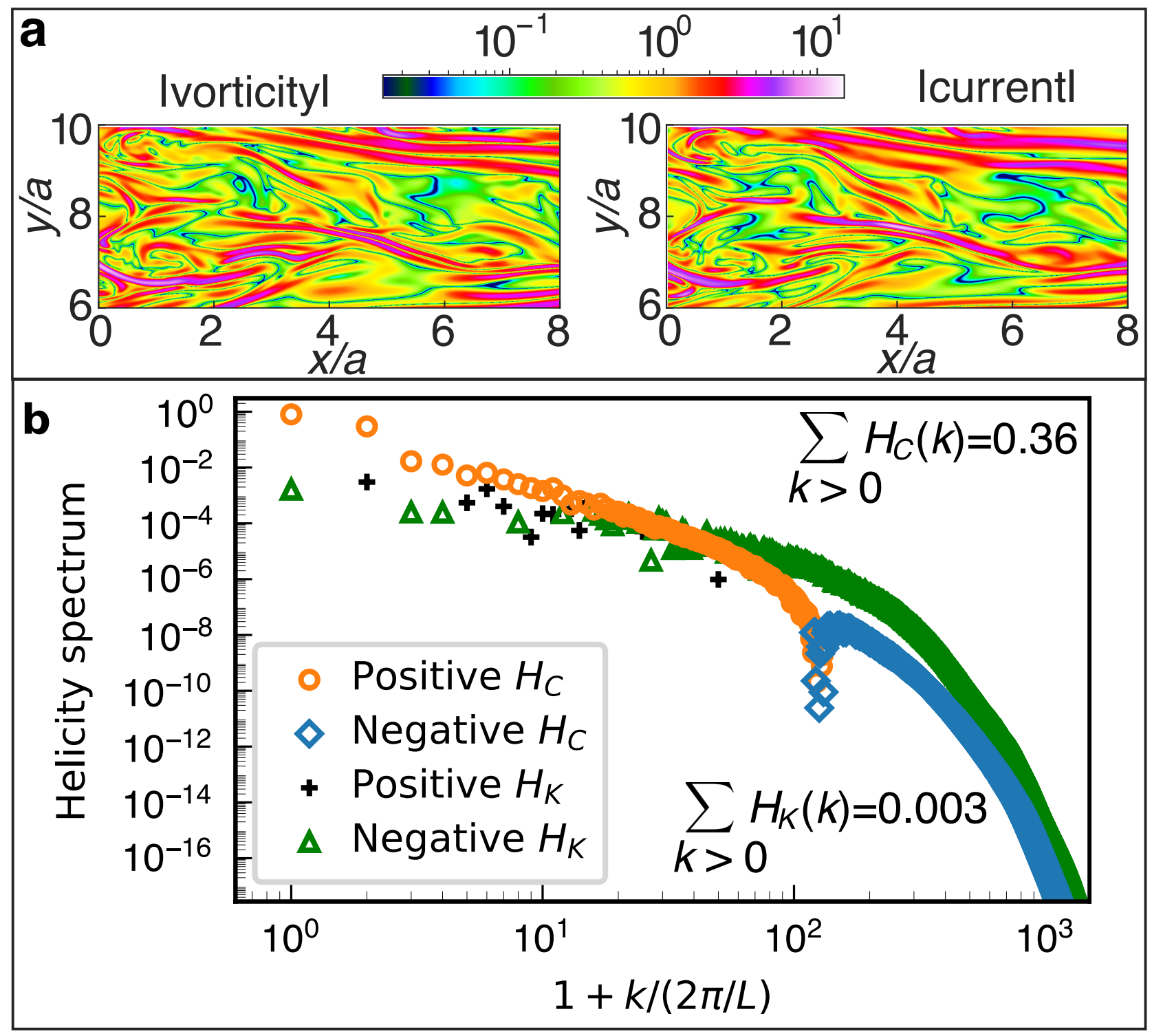

**Fig. 4 | Dominance of Cross-helicity. a**, Vorticity $|\omega_z|/\omega_z^{rms}$ and electric current density $|j_z|/j_z^{rms}$ are similar; rms stands for room mean square. A typical slice is shown at a shear-layer interface ($z$=3$L_z$/4); the full simulation domain is [0,10π)×[0,10π)×[0,20π). **b**, The total turbulent cross-helicity $H_C$ (= $\tilde{\mathbf{u}} \cdot \tilde{\mathbf{b}}$) is two orders of magnitude larger than the total turbulent kinetic helicity $H_K$ (= $\tilde{\mathbf{u}} \cdot \nabla\times\tilde{\mathbf{u}}$). Cross-helicity dominates at larger scales of fluctuations.

# Methods

**I. Model and computational considerations.**
Using 90 million Central Processing Unit (CPU) hours, we performed more than 90 KH dynamo simulations. We analyzed 0.25 petabytes of turbulence data generated by evolving velocity **u** and magnetic field **b** according to the 3D incompressible MHD equations

$$\partial_t \mathbf{u} - \nu\nabla^2\mathbf{u} + \nabla P = -\mathbf{u}\cdot\nabla\mathbf{u} + \frac{(\nabla\times\mathbf{b})\times\mathbf{b}}{\mu_0\rho} + \mathbf{f}, \quad \text{(S1a)}$$
$$\partial_t \mathbf{A} - \eta\nabla^2\mathbf{A} + \nabla\Psi = \mathbf{u}\times\mathbf{b}, \quad \text{(S1b)}$$
$$\mathbf{b} = \nabla\times\mathbf{A}, \quad \text{(S1c)}$$
$$\nabla\cdot\mathbf{u} = 0, \quad \text{(S1d)}$$
$$\nabla\cdot\mathbf{A} = 0, \quad \text{(S1e)}$$

where **A** is the magnetic vector potential; $\nu$ is the kinematic viscosity; $\eta$ is the electric resistivity; $\mu_0$ is the vacuum magnetic permeability; $P$ is the fluid pressure; $\Psi$ is a scalar function; and **f** represents external acceleration applied on a fluid of density $\rho$ (here, $\mu_0\rho=1$). Equation (S1c) ensures $\nabla\cdot\mathbf{b}=0$. We consider non-periodic and periodic domains. We simulate turbulence with and without an external mean magnetic field.

**II. Background profiles and non-dimensionalization.**
In non-periodic domains, we consider a single shear layer[10]

$$U_x(z) = U_0 \tanh\left(\frac{z}{a}\right), \quad \text{(S2)}$$

in a volume $[0,L)\times[0,L)\times[-L/2,L/2]$. In periodic domains, we consider a double shear layer[35]

$$U_x(z) = U_0\left[\tanh\left(\frac{z-z_1}{a}\right) - \tanh\left(\frac{z-z_2}{a}\right) - 1\right], \quad \text{(S3)}$$

in a volume $[0,L)\times[0,L)\times[0,2L)$. Here, $z_1=L/4$ and $z_2=3L/4$, with $L=10\pi$. Speed and length scales are measured using $U_0$ and $a$ throughout this work. Fluid and magnetic Reynolds numbers are $Re=U_0a/\nu$ and $Rm=U_0a/\eta$, respectively.

When a weak, uniform, external mean magnetic field $\mathbf{B}_0$ is initially imposed, $\mathbf{B}_0=(\hat{\mathbf{e}}_x\cos\theta+\hat{\mathbf{e}}_y \sin\theta)/M_A$, where $M_A$ is the Alfvénic Mach number, and θ is the angle between the initial mean field and the mean flow. Dynamo saturation properties remain largely unimpacted by variations in $M_A$ until $M_A$ approaches the KH-instability threshold. Variations in θ are found to not affect dynamo saturation.

We seed instability by adding broadband, phase-randomized, divergence-free perturbations[50] to **u** and **A**. The jet-driven dynamo emerges from general random perturbations. We solve Eqs. (S1a)–(S1e) using two disparate pseudospectral solvers: Dedalus (non-periodic and periodic domains) and GHOST (periodic domains).

**III. Dynamos in non-periodic and periodic domains.**

For non-periodic domains, we choose physical boundaries that confine the plasma within perfectly conducting ($j_x$=$j_y$=$b_z$=0), no-slip walls, represented by $u_z$=$u_y$=$A_x$=$A_y$=$\Psi$=0 and $u_x$=±$U_0$ at $z$=± $L$/2. When a uniform horizontal mean field $\mathbf{B}_0$ is present in the initial condition, we isolate that mean field from the remaining fields; the latter are subjected to the above magnetic boundary conditions.

The dynamo does not depend on the nature of the domain because it is enabled by large-scale jets, which are strongest near the shear layer. Both jets and shear layer lie maximally far away from the boundaries; see Sec. VIII for an analytic calculation. In some dynamos, one injects helicity fluxes through open, manipulated boundaries, hoping to achieve a large-scale dynamo. Here, the shear layer is a minimal and sufficient ingredient for exciting a dynamo. Our simulations with extended vertical domain reproduce the results reported here.

**IV. Benchmarks and major code optimizations.**

**Dedalus simulations**: Using Dedalus[51] (an open-source, general code written primarily in python), we employ MPI-parallelization along the periodic directions $x$ and $y$. We use pseudospectral $\tau$-method[35,51], RK443 and SBDF2 mixed implicit-explicit time-steppers[52,53], and the 3/2 dealiasing for nonlinearities[51]. For non-periodic domains, we use Fourier–Fourier–Chebyshev bases, whose harmonics range from $128^2$×512 to $1{,}024^3$.

**GHOST simulations**: Using GHOST[54,55] (an open-source code written primarily in Fortran 90/95 to efficiently solve triply periodic systems), we parallelize with MPI in $z$ and openMP in $y$. For efficient computations, we place OpenMP threads next to each other within a NUMA node and bind OpenMP threads to an MPI task. The nonlinear terms are solved using the second-order explicit Runge–Kutta scheme. We have benchmarked GHOST against Dedalus, using identical initial conditions[56]. We find the dynamo-enabling jets appear in all test and production simulations, whose resolutions we vary from $32^2$×512 to $4096^2$×8192. The high $z$-resolution captures the mean shear flow.

The public GHOST code solves the visco-resistive terms using an explicit method, which is not ideal to capture small-scale fluctuations. These fluctuations, if not resolved, can lead to numerical instability. We have implemented a new time-integrator, i.e., an exact integrating factor technique, which produces exact solutions corresponding to the visco-resistive terms. We have also implemented a forcing function, which maintains the mean flow.

**V. External mean-flow forcing.**

We externally force only the mean flow $u_x(k_x{=}0, k_y{=}0)$. The magnetic fields evolve freely. This allows us to focus on KH instability, which excites fluctuations at various scales via its nonlinear coupling. The forcing $\mathbf{f}=f_x(z)\hat{\mathbf{e}}_x$ in Eq. (S1a) is[57–62]

$$f_x(z) = \frac{\langle u_x(t=0)\rangle_{x,y} - \langle u_x(t)\rangle_{x,y}}{\tau_\mathrm{f}} + F_0, \qquad \text{(S5)}$$

where $\tau_\mathrm{f}$ is the forcing time scale; $F_0$ is added to merely remove the viscous relaxation of the mean shear flow at $t$=0, ensuring that the initial mean flow is a true MHD equilibrium, $F_0 + \nu\nabla^2\langle u_x(t=0)\rangle_{x,y}=0$.

We vary $\tau_f$ from 0 (frozen mean flow) to infinity (decaying mean flow). We reproduce essentially the same results except when $\tau_f$ is very large. In the latter case, the gradient of the mean shear flow quickly flattens because its energy is extracted by the KH instability at a rate faster than the rate of energy injection to the mean flow by the external forcing.

**VI. Analytic quasilinear EMF model.**
Consider an ($x$,$y$)-averaged mean magnetic field **B**=($B_x$, $B_y$, 0) and mean flow **U**=($U_x$, $U_y$, 0). We introduce external perturbations in flow **u**'=(0,0,$u_z$') and fields **b**'=(0,0,$b_z$') with wavenumber **k**=(0,$k_y$); primed quantities in this subsection represent perturbations. This wavenumber dominantly generates mean magnetic field (Extended Data Fig. 3). In a quasilinear model, $u_z$' and $b_z$' interact with the mean flow and mean field to self-consistently generate other perturbations, e.g., $u_x$' and $b_x$'. These, together with the initial perturbations, compose a mean turbulent electromotive force, which causes the mean magnetic field to evolve.

The ($x$,$y$)-averaged mean magnetic field $B_x$ evolves according to $(\partial_t - \eta\partial_z^2)B_x = -\partial_z\mathcal{E}_y$, where

$$\mathcal{E}_y = -b_z'^* u_x' + u_z'^* b_x' + c.c., \qquad \text{(S6)}$$

where *c.c.* represents complex conjugation of the preceding terms.

To derive expressions for $u_x$' and $b_x$' in terms of the initial small-amplitude perturbations $u_z$' and $b_z$', we write

$$\left(\partial_t - \nu\nabla^2 + U_y i k_y\right)u_x' = -u_z'\,\partial_z U_x + b_z'\,\partial_z B_x + B_y i k_y b_x', \qquad \text{(S7)}$$

and

$$\left(\partial_t - \eta\nabla^2 + U_y i k_y\right)b_x' = -u_z'\,\partial_z B_x + b_z'\,\partial_z U_x + B_y i k_y u_x'. \qquad \text{(S8)}$$

The solutions to $u_x$' and $b_x$' are approximately written as

$$u_x' = \tau_{\mathrm{ZF}}\left[-u_z'\,\partial_z U_x + b_z'\,\partial_z B_x + B_y i k_y b_x'\right], \qquad \text{(S9)}$$

and

$$b_x' = \tau_{\mathrm{ZM}}\left[-u_z'\,\partial_z B_x + b_z'\,\partial_z U_x + B_y i k_y u_x'\right], \qquad \text{(S10)}$$

where $\tau_{\mathrm{ZF}}$ and $\tau_{\mathrm{ZM}}$ are the coherent-straining times of the zonal jet $u_x$' and zonal magnetic field $b_x$'. Here, the word *zonal* refers to *x-averaged fluctuations.* When the effects of the nonlinear turbulent interactions are considered, $\tau_{\mathrm{ZF}}$ and $\tau_{\mathrm{ZM}}$ are replaced by nonlinearly modified times, which allows the calculations to be more generally applicable than merely in the small-amplitude limit.

Substituting the solutions for the zonal jet [Eq. (S9)] and zonal field [Eq. (S10)] in Eq. (S6),

$$\mathcal{E}_y = \Upsilon(\nabla \times \mathbf{U})_y - \beta(\nabla \times \mathbf{B})_y + \alpha B_y, \quad \text{(S11)}$$

where

$$\Upsilon = \underbrace{\tau_{\mathrm{ZM}}\langle u_z' b_z' \rangle_{x,y}}_{\text{kinematic}} + \underbrace{\tau_{\mathrm{ZF}}\langle u_z' b_z' \rangle_{x,y}}_{\text{non-kinematic}}, \quad \text{(S12a)}$$

$$\beta = \underbrace{\tau_{\mathrm{ZM}}\langle u_z'^2 \rangle_{x,y}}_{\text{kinematic}} + \underbrace{\tau_{\mathrm{ZF}}\langle b_z'^2 \rangle_{x,y}}_{\text{non-kinematic}}, \quad \text{(S12b)}$$

$$\alpha = \underbrace{-\tau_{\mathrm{ZM}}\langle u_z' \omega_z' \rangle_{x,y}}_{\text{kinematic}} + \underbrace{\tau_{\mathrm{ZF}}\langle b_z' j_z' \rangle_{x,y}}_{\text{non-kinematic}}. \quad \text{(S12c)}$$

Equations (S11) and (S12a)–(S12c) were first derived in analogous forms using a turbulence closure[4] (see also Sec. 2 of Ref. 46).

The non-kinematic terms appearing with $\tau_{\mathrm{ZF}}$ in Eq. (S12a)–(S12c) arise from the time-evolution of velocity fluctuation in Eqs. (S7) and (S9). Generally, $\tau_{\mathrm{ZF}} = \tau_{\mathrm{ZM}}$. Thus, for pure Alfvénic states ($\mathbf{u}||\pm\mathbf{b}$), $\alpha$ reduces to zero[28,63], whereas $\beta$ and $\Upsilon$ become maximally non-zero.

**VII. Zonal jets in the KH dynamo.**

Energetically dominant, large-scale jets are generated when fluid motion $u_z(k_x=0,k_y)$ is stretched by the mean flow (Fig. 3**a**). The motion $u_z(k_x=0,k_y)$ is *nonlinearly* generated by the KH instability, as the vertical mean flow and vertical mean field are absent; the KH instability exists only at $k_x \neq 0$ (Extended Data Fig. 1).

The motion $u_z(k_x \neq 0,k_y)$, on the other hand, is sheared by the mean flow on the $(x,z)$-plane, thus progressively being shortened in its vertical scale until it is broken into multiple fragments[64]; see Fig. 1 of Ref. 3. However, this two-dimensional fragmentation does not apply to the three-dimensional *x-invariant* fluctuation, which can be shown using rapid distortion theory[65]. The mean flow maintains the $z$-spatial scale of $u_z(k_x=0,k_y)$ and produces strong jets $u_x(k_x=0,k_y)$. The jets are minimally impacted by the Lorentz feedback because the forces on them from the gradients of fluid pressure and magnetic pressure are zero. The jets are robust (Fig. 2**a**).

**VIII. Analytic semi-local dynamo characterization.**

To characterize the generation of the mean magnetic field that is reversed across the shear layer $z$=0, we expand around $z$=0 the mean flow $u_x^{0,0}$ [the superscript represents the horizontal wavenumber $(k_x,k_y)$]

$$u_x^{0,0} = \tanh(z) = z \underbrace{-z^3/3 + \cdots}_{\text{KH dynamo source}}, \quad \text{(S13)}$$

where the terms nonlinear in $z$ are nonlocal. We focus on $z^3$ in the semi-local analysis below.

Energy transfer analysis informs us that jets are formed when the mean flow stretches $u_z^{0,k_y}$. Hence, we write

$$\partial_t u_x^{0,k_y} = -\mathbf{u}^{0,k_y} \cdot \nabla u_x^{0,0} + \cdots = -u_z^{0,k_y}\,\partial_z u_x^{0,0} + \cdots = -u_z^{0,k_y}(1 - z^2 + \cdots) + \cdots, \qquad \text{(S14)}$$

where … represents terms not relevant here. The solution of Eq. (S14) is approximately written as

$$u_x^{0,k_y} = -\tau_c u_z^{0,k_y}\left(1 - \underbrace{z^2 + \cdots}_{\text{jets symmetric in } z}\right) + \cdots, \qquad \text{(S15)}$$

where $\tau_c$ is a time scale of flow-straining that generates $u_x^{0,k_y}$. Equation (S15) shows that the jet amplitude $u_x^{0,k_y}$ is the strongest at the shear-layer interface $z$=0 and decreases symmetrically with increasing $|z|$. This can be seen in Figs. 1**c** and 1**d**.

Jets $u_x(k_x=0,k_y)$ generate the mean field (Extended Data Fig. 3) via the field-line stretching term $\mathbf{b}^{0,-k_y}\cdot\nabla u_x^{0,k_y}=\nabla_j\,(b_j^{0,-k_y}u_x^{0,k_y})$, where the repeated index $j$ follows the Einstein's summation convention. We simplify this term by using Eq. (S15) to find

$$\partial_t b_x^{0,0} = \partial_z\left(b_z^{0,-k_y}u_x^{0,k_y}\right) + \cdots = -\tau_c b_z^{0,-k_y}u_z^{0,k_y}\,\partial_z[1 - z^2 + \cdots] + \cdots$$
$$= \underbrace{2z}_{\text{mean field reversed in } z} \times \tau_c u_z^{0,k_y} b_z^{0,-k_y} \ldots + \cdots, \qquad \text{(S16)}$$

where … represents terms not relevant here.

The linear term $z$ in Eq. (S16) emerges from $\partial_z^2 u_x^{0,0}$ and makes the mean magnetic field $b_x^{0,0}$ reversed across $z$=0 (Fig. 1**a**). The polarity of the mean field flips over time due to the phase difference between $u_z^{0,k_y}$ and $b_z^{0,-k_y}$**.** This correlation measures the turbulent cross-helicity ϒ. Equation (S16) suggests

$$b_x{}^{0,0} \propto \hat{\mathbf{e}}_x \cdot [\nabla\times(\Upsilon\nabla\times\mathbf{u}^{0,0})]. \qquad \text{(S17)}$$

The foregoing analyses reveal that the Yoshizawa's postulated generic ϒ-effect[4] operates specifically via the formation of 3D jets (see also Fig. 3).

**IX. Spectral energy-flux computations.**

The KH turbulence is strongly anisotropic and localized around the shear layer. So, for post-processing, it is not ideal to use spherical wavenumber shells. We employ cylindrical shells in the $(k_x,k_y)$-plane with the axis of the cylinder along $z$.

The $u$-to-$u$ energy flux $\Pi_{uu}(k_0)$ passing through the wavenumber $k_0$ is[66]

$$\Pi_{uu}(k_0) = \frac{1}{L_x L_y L_z}\int_x\int_y\int_z \mathbf{u}^{>} \cdot [-(\mathbf{u}\cdot\nabla)\mathbf{u}]\,dxdydz, \qquad \text{(S18)}$$

where **u** represents the entire velocity in the system. The velocity $\mathbf{u}^{>}$ is[66]

$$\mathbf{u}^{>} = \sum_{k_x,k_y:\left(k_x^2+k_y^2\right)^{1/2}>k_0} \hat{\mathbf{u}}\left(k_x,k_y,z\right)\mathrm{e}^{ik_x x+ik_y y}, \quad \text{(S19)}$$

where $\hat{\mathbf{u}}\left(k_x,k_y,z\right)$ is the Fourier amplitude of the velocity with wavenumber ($k_x$, $k_y$).

Similarly, we define the *b*-to-*b* energy flux $\Pi_{bb}(k_0)$ passing through the wavenumber $k_0$ using[66]

$$\Pi_{bb}(k_0) = \frac{1}{L_x L_y L_z}\int_x\int_y\int_z \mathbf{b}^{>}\cdot[-(\mathbf{u}\cdot\nabla)\mathbf{b}]dxdydz, \qquad \text{(S20)}$$

where **b** represents the entire magnetic field in the system. The magnetic field $\mathbf{b}^{>}$ is defined similar to $\mathbf{u}^{>}$ in Eq. (S19). Computations of these energy fluxes show cascades of kinetic and magnetic energies from large to small physical scales (Extended Data Fig. 2).

**X. Relevance of the ϒ-dynamo to laboratory plasmas.**

In addition to the fifth-to-last paragraph in the main article and Extended Data Fig. 4, we provide here details comparing the ϒ-dynamo with the Madison Dynamo Experiment[37] (MDE). While there are some differences between the present system and the MDE (such as spherical vs. Cartesian geometry), there are also similarities:

1. Both systems exhibit large-scale vortical flows and jets. The radial vorticity in the MDE is considerably large[38], consistent with the dominant radial EMF (Extended Data Fig. 4**b**).
2. The large-scale flows in both systems are externally maintained: in MDE by rotating impellers and in our simulations by the mean-flow forcing.
3. Both systems exhibit identical alignment of the EMF components. The angle between the mean magnetic field and EMF[37] is ~90°, confirming the $\alpha$-effect is virtually zero.
4. Both systems show orientations of the EMF components do not depend on microphysical dissipation properties[37].
5. Both systems find the dynamo is driven by large-scale energy-containing flows. Thus, the dynamo growth rate, as measured in simulations of the laboratory MDE[67], is asymptotically independent of the magnetic Reynolds number *Rm*. Compare this with a dynamo driven by near-viscous-scale motions, where the growth rate scales as $Rm^{1/2}$.

These findings suggest that the jet-driven dynamos are present in laboratory and nature.

**XI. Relevance of the ϒ-dynamo to astrophysical plasmas and BNS mergers.**

We address here whether the reported dynamo mechanism is relevant in astrophysical plasmas. The shear-flow-driven dynamo exists in a wide variety of astrophysical systems. We shall therefore present only two specific cases.

*Binary Neutron Star (BNS) Mergers*: Global GRMHD simulations of BNS mergers show the flow profile at the merger interface is similar to the flow profile this work considers. The shear layer half-width is detected[5] to be $a \approx 10$–15 m. Since this shear layer appears at small scales of global simulations, the flow velocity is around $U_0 \lesssim 0.1c$. Within the shear layer, the speed is even smaller. While the spacetime curvature effects are important at the global scale, they are very weak and are amenable to being transformed away in this extremely thin shear layer. This was exploited by Price and Rosswog[7], who used Newtonian KH simulations of BNS mergers. While compressibility introduces adjustments to the KH growth rates, studies of compressible KH systems have shown similar dynamics of the instability as observed in incompressible systems[7,68]. Density variation in the shear layer is not large. These justify the incompressible model for the first study of the reported new dynamo mechanism.

Major current challenges of BNS merger simulations include the use of unrealistically large-amplitude initial magnetic field and the choice of initial field topology (dipolar fields of $\sim 10^{13}$–$10^{15}$ G are often considered)[6,7,69,70]. However, neither the amplitude nor the topology is supported by astrophysical observations. Observations of Gyr-old neutron stars[71,72] show that their initial total magnetic fields are on the order of $<10^{8}$–$10^{10}$ G. Initial topology of realistic BNS magnetic fields is multipolar and complex. There are some debates on whether realistic magnetic fields that are initially weak can actually be amplified to magnetar-strength fields. Global relativistic simulations do not capture all KH-unstable wavenumbers appearing at smaller scales in their resolution-limited simulations[5]. They produce increasingly strong magnetic fields with increased grid-resolution.

It is in the context of the preceding paragraph that the results of this work find significance. The results show that, even when the initial large-scale magnetic field is zero, the KH instability generates strong large-scale fields in milliseconds from infinitesimal magnetic fluctuations. Such strong fields are then advected outward to cover the surface of the remnant, impacting observables such as electromagnetic emissions, detected by the LIGO-Virgo-KAGRA collaboration[42]. A proper modeling of BNS merger requires a detailed understanding of the KH dynamo, as the generated magnetic fields can significantly impact the merger time and radiated gravitational waves[8,73]. These fields are potentially responsible for the post-merger evolution of the resultant compact object via efficient transport of angular momentum by turbulence. Other observable consequences include the formation of extraordinarily powerful jets, which depend on the topology and strength of the magnetic fields formed during the BNS mergers. This work provides a foundational framework that can be used to capture the dynamo effects of unresolved KH fluctuations in global simulations[44,74].

*The Solar Magnetism*: The large-scale poloidal field generation mechanism in the Sun is modeled using ad-hoc, fine-tuned profile of the $\alpha$-coefficient[75]. The present work suggests a solar dynamo model where the $\alpha$-effect is replaced with the $\Upsilon$-effect arising from the azimuthal vorticity of the large-scale solar meridional circulation[14,15]. This azimuthal vorticity contributes to the azimuthal EMF ($\boldsymbol{\mathcal{E}} \propto \Upsilon \nabla \times \mathbf{U}$), which should generate poloidal magnetic fields. Additionally, helioseismology informs that solar sub-surface flows and near-surface shear layers possess substantial vorticity. Cross helicity has been measured at the solar surface using the data from the Hinode satellite and the Swedish 1-m Solar Telescope[76]. Numerical simulations indicate that the normalized cross-helicity is closer to unity near the solar surface[46]. Thus, the near-surface shear layers[12,77] are the potential sites for the generation of magnetic fields via the vorticity-induced $\Upsilon$-dynamo.

**XII. How does the jet-driven ϒ-dynamo differ from the classical and other dynamos?**
Perhaps the most popular dynamo is the $\alpha^2$-dynamo[20]. A large-scale magnetic field $B_x$ evolves as $(\partial_t\text{-}\eta\nabla^2)B_x=\hat{\mathbf{e}}_x\cdot(\nabla\times\boldsymbol{\mathcal{E}})$. The mean turbulent EMF $\boldsymbol{\mathcal{E}}$ is $\boldsymbol{\mathcal{E}}=\alpha\mathbf{B}\text{-}\beta\nabla\times\mathbf{B}$ for helical turbulence[24]. Since $B_x$ couples to another component $B_y$ (or $B_z$) of the large-scale magnetic field, a way to close the dynamo loop is to write $(\partial_t\text{-}\eta\nabla^2)B_y=\hat{\mathbf{e}}_y\cdot(\nabla\times\boldsymbol{\mathcal{E}})$ and assume, again, $\boldsymbol{\mathcal{E}}=\alpha\mathbf{B}\text{-}\beta\nabla\times\mathbf{B}$. One of the two steps of the $\alpha^2$-dynamo is replaced by the Ω-effect if there exists a large-scale flow whose gradient (magnitude given by Ω) points in the direction of the mean magnetic field. This $\alpha\Omega$-dynamo has been numerically observed to be replaced by an $A\Omega$-dynamo in a case of a linear shear flow, where $A$ is to-this-date an unconfirmed mechanism existing in their simulation[78]. A variant of this dynamo is where $A$ is replaced with an anisotropic[23] turbulent diffusion $\beta_{ij}$. These and other dynamos share a common feature: The two steps of the dynamo couple two components of large-scale magnetic fields, because all these dynamo processes assume that $\boldsymbol{\mathcal{E}}$ depends on $\mathbf{B}$ and its spatial properties.

We highlight how the jet-driven ϒ-dynamo is fundamentally different from the classical and other dynamos. When a large-scale shear flow $\mathbf{U}$ is present, the postulated generalized mean EMF[4] is $\boldsymbol{\mathcal{E}}=\Upsilon\nabla\times\mathbf{U}\text{-}\beta\nabla\times\mathbf{B}$, where $\Upsilon\propto\langle\tilde{\mathbf{u}}\cdot\tilde{\mathbf{b}}\rangle$ measures the alignment between turbulent flow $\tilde{\mathbf{u}}$ and turbulent magnetic field $\tilde{\mathbf{b}}$. Hence, the mean magnetic field $B_x$ along the direction of the mean flow $U_x$ evolves as

$$[\partial_t - (\eta+\beta)\nabla^2]B_x=\hat{\mathbf{e}}_x\cdot[\nabla\times(\Upsilon\nabla\times U_x)]. \qquad \text{(S28)}$$

It is *not* necessary to write the evolution equation of the passive components $B_y$ (or $B_z$) because Eq. (S28) couples only to $U_x$, $\beta$, and ϒ. Since the traditional mean-field kinematic dynamo coefficients (e.g., $\alpha$, $\beta$, $A$, $\beta_{ij}$) are justifiably constants in time, it is tempting to assume that ϒ is also a constant. This suggests $B_x$ grows only linearly in time[79,80]. Upon inspection, we realize that $\beta$ (similar to $\alpha$) depends on turbulent flow only, but ϒ depends also on the magnetic field fluctuations. Since the mean and fluctuations of magnetic fields are inevitably coupled, and both grow at the same rate in the kinematic phase—as we have observed in simulations—ϒ grows exponentially in time.

To obtain the correct growth of $B_x$ in Eq. (S28), we analytically derive a self-consistent evolution equation for ϒ, which measures cross-helicity in fluctuations. We find that the volume-averaged turbulent cross-helicity ϒ evolves as

$$(\partial_t + c)\Upsilon = f(B_x), \qquad \text{(S29)}$$

where $c$ is a constant, and $f(B_x)$ is a linear operator that integrates over volume the mean field $B_x$ with a weight factor, which depends on the mean flow $U_x$. Thus, Eqs. (S28) and (S29) are linearly coupled equations and produce exponential solutions. Because of the inhomogeneity in the mean flow, the solutions require calculations in Fourier space. Since the calculations are lengthy and technical, we will report the details of the calculations in a forthcoming publication.

This work also presents an interpretation, visualization, and comparison of the non-traditional ϒ-dynamo with the traditional dynamo; see Fig. 3**a**. We expect the ϒ-effect to drive the dynamo of Ref. 81. The ϒ-effect may exist in the dynamo of Ref. 82, as well.

*Differences between the ϒ- and Ω-effects*: The ϒ-effect arises in a mean-field theory where the interactions between turbulent fluctuations are averaged. The Ω-effect does not require a mean-field theory and simply represents the straining of the large-scale magnetic field **B** by the large-scale flow **U**. The large-scale field **B**, in general, evolves as

$$\partial_t \mathbf{B} = \nabla \times [(\mathbf{U} \times \mathbf{B}) + \boldsymbol{\mathcal{E}}] + \eta \nabla^2 \mathbf{B}, \qquad \text{(S30)}$$

where $\boldsymbol{\mathcal{E}}=\langle \tilde{\mathbf{u}} \times \tilde{\mathbf{b}} \rangle$ is the turbulent electromotive force averaged suitably over some scales. Using a generalized mean-field theory $\boldsymbol{\mathcal{E}}=\alpha\mathbf{B}-\beta\nabla\times\mathbf{B}+\Upsilon\nabla\times\mathbf{U}$, we simplify Eq. (S30) to

$$\partial_t \mathbf{B} = \underbrace{-\Upsilon\nabla^2\mathbf{U}}_{\substack{\Upsilon\text{-effect} \\ \text{(source)}}} + \underbrace{\beta\nabla^2\mathbf{B}}_{\substack{\beta\text{-effect} \\ \text{(sink)}}} + \underbrace{\alpha\nabla\times\mathbf{B}}_{\substack{\alpha\text{-effect} \\ (\approx 0)}} + \underbrace{\mathbf{B}\cdot\nabla\mathbf{U}}_{\substack{\Omega\text{-effect} \\ =0}} + \cdots, \qquad \text{(S31)}$$

where … represents terms not relevant here. How the KH dynamo operates is indicated by the text below each term.

The jet-driven ϒ-dynamo presented in this work is radically different from the classical and other dynamos in many fundamental ways as summarized above.

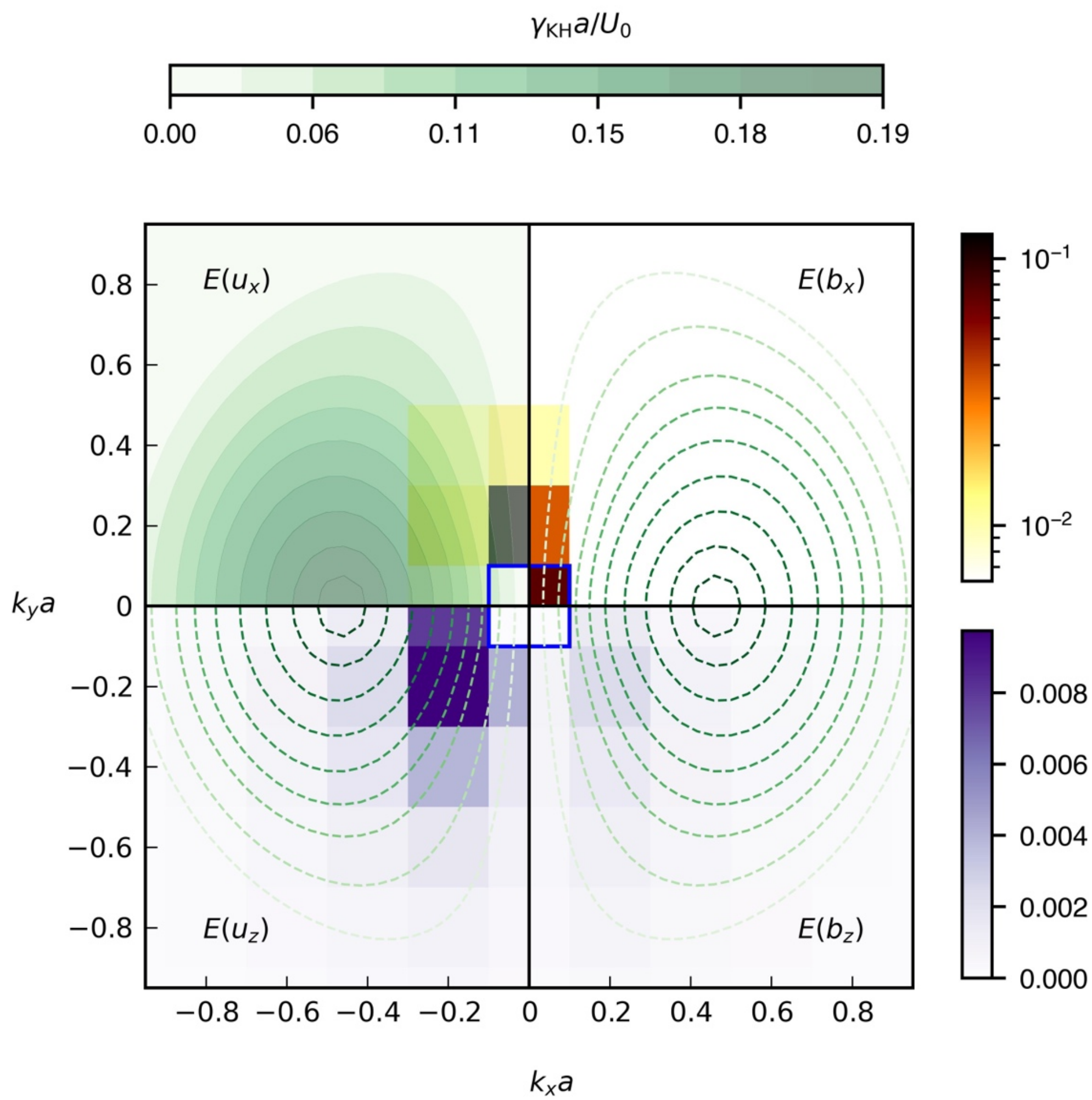

**Extended Data Fig. 1 | Anisotropic growth rate spectrum of the KH instability, along with the spectra of energies.** The mean flow is directed along $x$ and varies along $z$. The spectrum of the linear growth rate $\gamma_{KH}$ of the KH instability is anisotropic. The growth rate is the largest around $\mathbf{k}$=(±0.5, 0), which represents the 2D KH instability. The growth rate is zero for $k_x$=0 and $k_y$≠0. The square boxes in the upper half of the figure show time-averaged spectra of energies in horizontal field $E(b_x)$ and horizontal flow $E(u_x)$ (the contribution of the initial mean flow is removed). These spectra share the upper right-hand logarithmic color bar, extending from yellow to black. The square boxes in the lower half of the figure show time-averaged spectra of energies in vertical field $E(b_z)$ and vertical flow $E(u_z)$. These share the lower right-hand linear color bar, extending from white to blue. The center of each square box represents the wavenumber resolved in the nonlinear simulation (only a small part of the wavenumber range is shown, as the KH instability exists at large scales $|ka|\lesssim 1$). The special square box centered at $\mathbf{k}=\mathbf{0}$ is shared by four quadrants. All quantities are measured using $a$ and $U_0$.

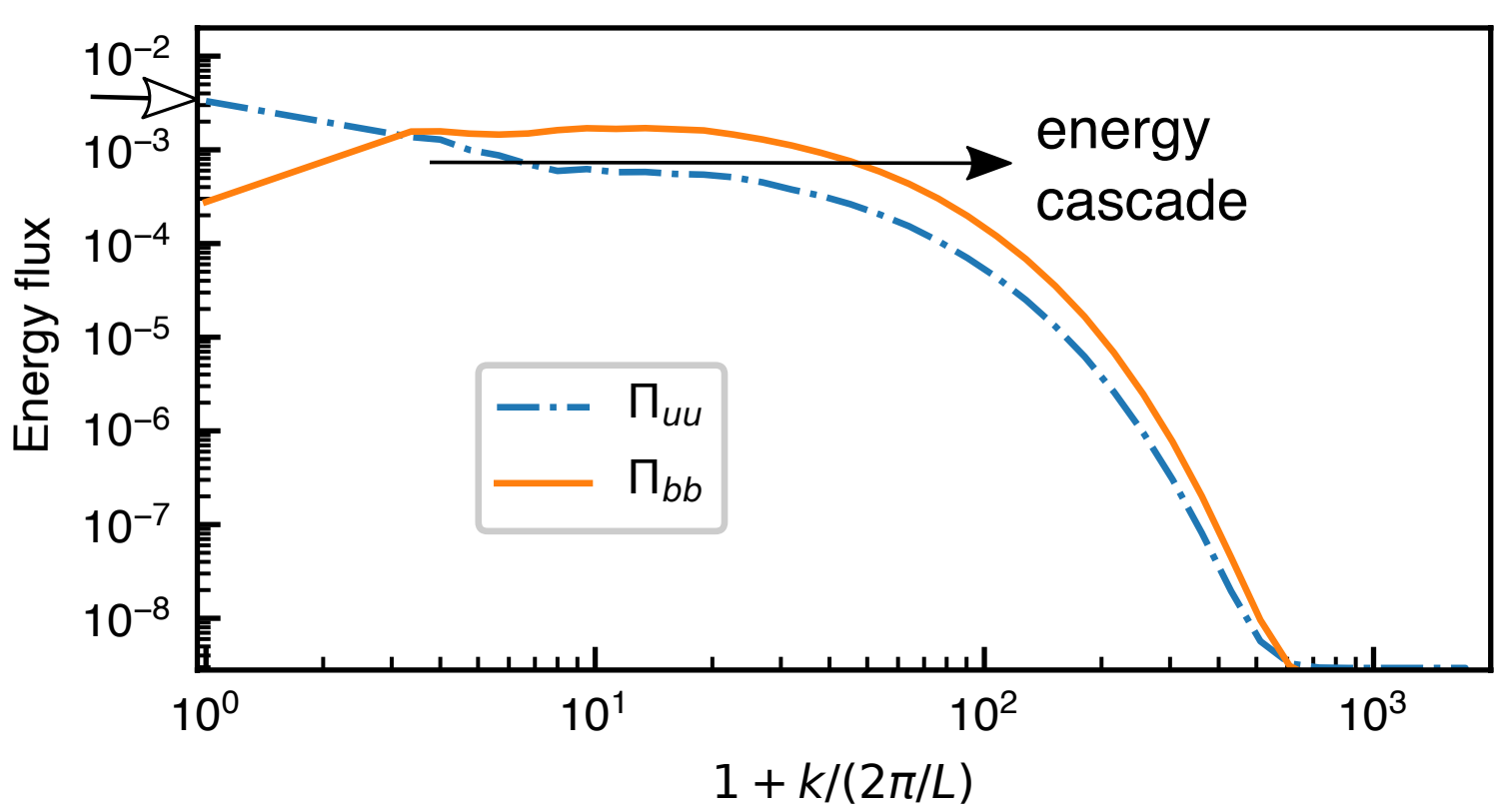

**Extended Data Fig. 2 | Cascade of kinetic and magnetic energies from large scales to small scales.** Nonlinear energy flux through spectral space is measured in a simulation with 4096 × 4096 × 8192 grid points. Here, $k=(k_x^2+k_y^2)^{1/2}$. The energy flux is integrated over the $z$-axis. A constant energy flux in $k$-space indicates an inertial range. Energy injected externally to the mean flow (shown with a white-headed arrow on the left margin) is cascaded to small scales, as shown by $\Pi_{uu}$, which measures energy flux due to transfer of energy between two velocity fluctuations. Magnetic-to-magnetic energy flux $\Pi_{bb}$ shows a prominent small-scale cascade over two decades of inertial range. This analysis shows that the Υ-dynamo arising in KH-unstable shear flow is *not* due to inverse cascade of energy. The Υ-dynamo is due to energy transfer from large-scale jets to the mean field, as demonstrated in Extended data Fig. 3.

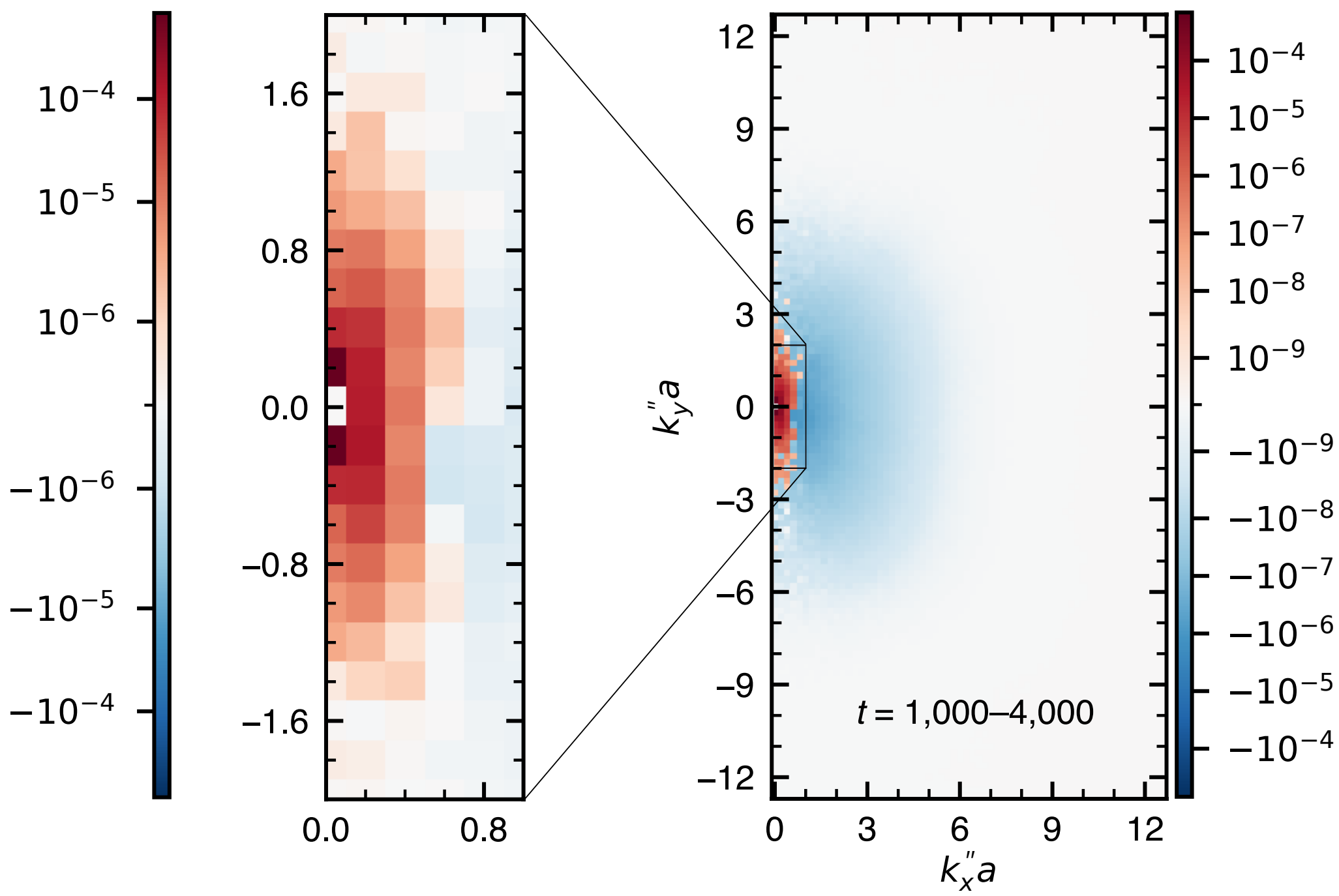

**Extended Data Fig. 3 | Spectrum of nonlinear energy transfer to the mean magnetic field via turbulent field-line stretching.** This figure confirms that the large-scale velocity fluctuations give energy to the mean magnetic field (red)—and that the small-scale velocity fluctuations receive energy from the mean magnetic field (blue). The shown quantity is $T(\mathbf{k}'')$, which represents the rate of nonlinear energy transfer to the mean field $\hat{b}_x(k_x=0, k_y=0)$. Mathematically, $T(\mathbf{k}'') = \langle \hat{b}_x^*(0,0)[\hat{\mathbf{b}}(\mathbf{k}')\cdot\nabla\hat{u}_x(\mathbf{k}'')]\rangle_z$, where $\mathbf{k}''=(k_x'',k_y'')$ is the horizontal wavenumber of the velocity $\mathbf{u}$ and $\mathbf{k}'=(k_x',k_y')$ is the horizontal wavenumber of the magnetic field $\mathbf{b}$; the wavenumber-triad constraint imposes $(0,0) = \mathbf{k}' + \mathbf{k}''$. The quantities $\hat{\mathbf{b}}(\mathbf{k}')$ and $\hat{u}_x(\mathbf{k}'')$ represent the horizontally Fourier-transformed coefficients of magnetic fields and velocity with wavenumbers $\mathbf{k}'$ and $\mathbf{k}''$, respectively; these quantities are retained in the physical domain $z$ before integrating the transfer function along $z$. The quantity $\hat{b}_x^*(0,0)$ is the horizontally-Fourier-transformed, complex-conjugated, $x$-directed mean magnetic field, whose energy is around two orders of magnitude larger than the $y$-directed mean field in the nonlinear phase. The mean field $\hat{b}_x(0,0)$ is inhomogeneous in $z$ (as is the mean flow); the operation $\langle\cdot\rangle_z$ averages the nonlinear transfer function in $z$. The transfer function is time-averaged over the saturated phase ($t$ = 1000–4000). Of particular note is the extraordinary contribution of the large-scale jets, especially with wavenumber $(k_x'',k_y'')=(0,0.2)$, which contribute the largest in the mean-field generation. This is consistent with Fig. 3. The square box at $k_x''=k_y''=0$ is white because the mean flow does not directly couple to the mean field in this system (the so-called Ω-effect[32] is zero).

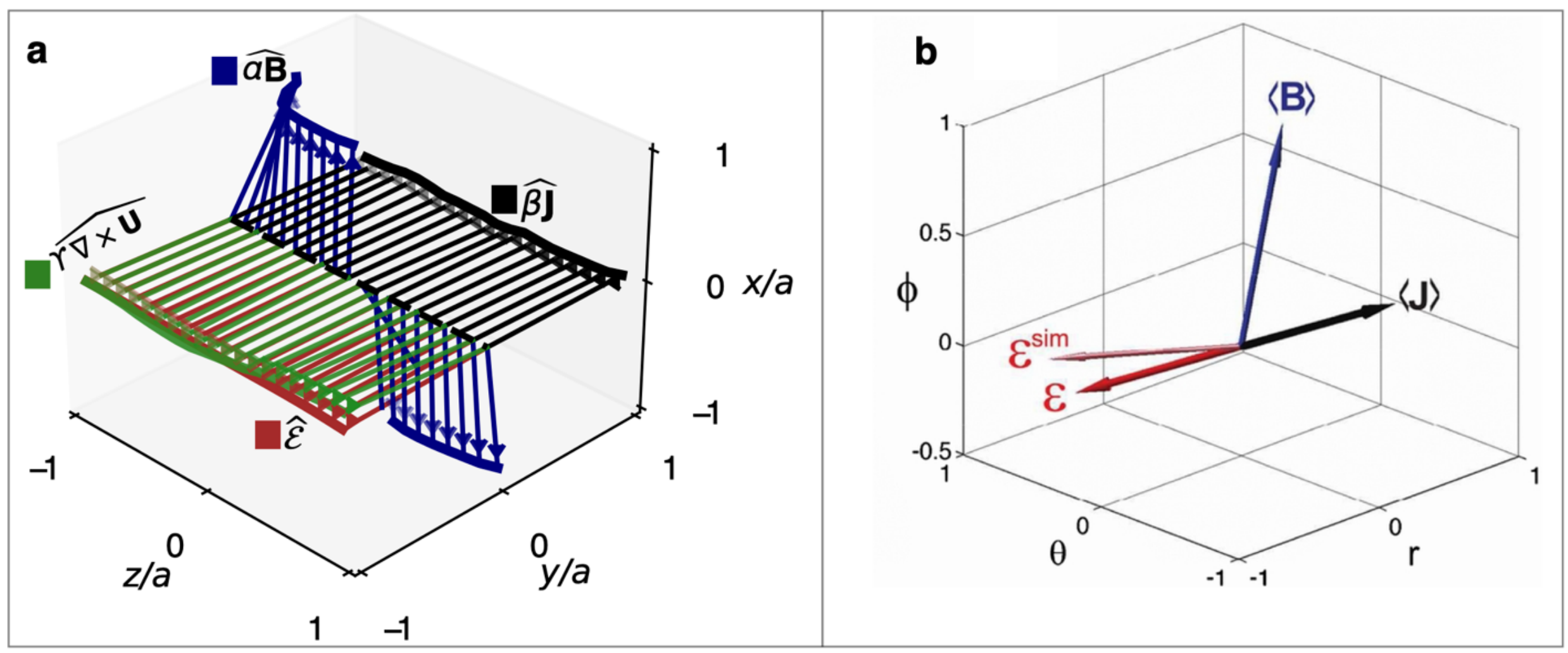

**Extended Data Fig. 4 | Comparison of three-dimensional unit vectors (shown with carets) of different components of the mean turbulent EMF $\boldsymbol{\mathcal{E}}$ from a KH-instability-driven dynamo simulation in a and from the Madison Dynamo Laboratory Experiment[37] in b.** With respect to $\boldsymbol{\mathcal{E}}$, the $\beta$-diffusion term is anti-aligned, and the $\alpha$-term is orthogonal (indicating the non-helical nature of dynamo). A near-identical orthogonal orientation of the $\alpha$-term was measured in the Madison Dynamo Experiment[37] in panel **b**. In the Madison Dynamo Experiment, the vorticity is considerably large in the radial direction[38]. This is consistent with the dominance of the observed radial EMF, suggesting the important role of the large-scale vorticity-effect[4] in turbulent EMF. Similarly, the ϒ-term in panel **a** is perfectly co-aligned with the mean EMF $\boldsymbol{\mathcal{E}}$. Here, the novel jet-driven ϒ-dynamo, arising from the large-scale vorticity, is confirmed to be the source of the dynamo. Diagram in **b** reproduced with permission of the AAS from figure 3 in ref. 37.

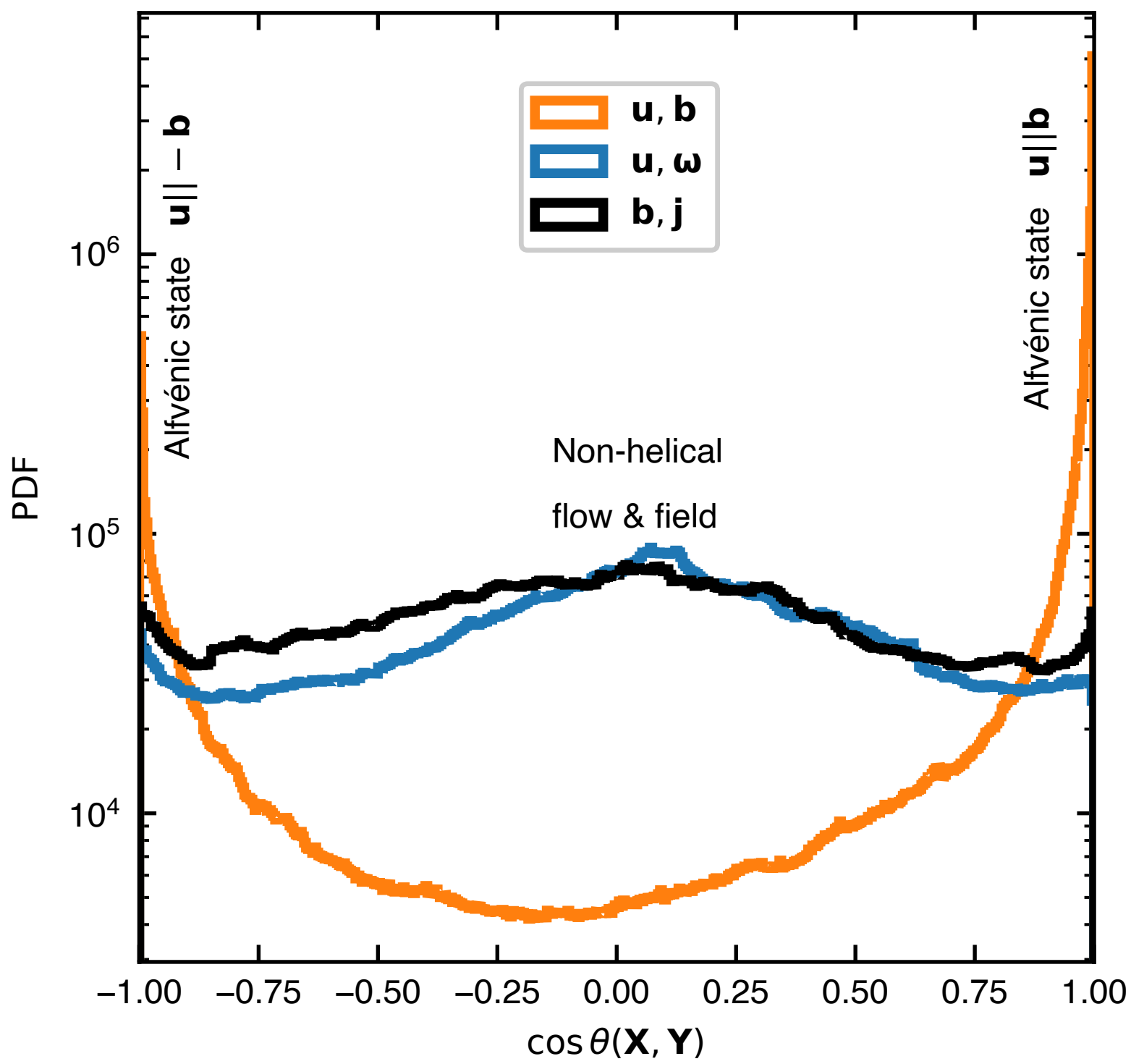

**Extended Data Fig. 5 | PDF of cosine of angle between u, b, ω(=∇×u), and j(=∇×b), in a simulation with 4096 × 4096 × 8192 grid points.** The turbulent flow and fields measured at the shear layer are non-helical and close to pure Alfvénic states (orange curve), which explains why the ϒ-effect dominates in the mean EMF. For the pure Alfvénic states, $\alpha$ is zero; $\beta$ and ϒ are similar and are not impacted by the non-kinematic (flow-evolution) effect in the same manner as the $\alpha$ is (see Methods Sec. VI). A large-scale steady shear flow induces an "imbalanced" MHD turbulence (asymmetry of the orange curve) and thus drives the dynamo via the non-zero ϒ-coefficient. Similar PDFs of dominant cross-helicity and non-helical flows were recently detected in the magnetosheath turbulence observed by the Magnetospheric Multiscale Spacecraft[49].

**Extended Data Table 1 | A list of commonly used symbols and meanings.**

| Symbols | Meanings |
|---|---|
| $\langle\cdot\rangle_{x,y}$ | $(x,y)$-average, also called *mean* |
| $\langle\cdot\rangle_{x}$ | $x$-average, also called *zonal* |
| $\langle\cdot\rangle_{y}$ | $y$-average |
| $\mathbf{u}_0$ | Initial mean flow |
| $\mathbf{b}_0$ | Initial mean magnetic field |
| $\mathbf{U}$ | Instantaneous mean flow |
| $\mathbf{B}$ | Instantaneous mean magnetic field |
| $\tilde{\mathbf{u}}$ | Flow fluctuations (i.e., without the mean) |
| $\tilde{\mathbf{b}}$ | Magnetic-field fluctuations (i.e., without the mean) |
| $U_0$ | Amplitude of the mean flow |
| $a$ | Half-width of the mean flow (shear layer) |
| $\tau_{\text{grow}}$ | One e-folding time of the KH instability |
| $Re$ | Fluid Reynolds number ($Re=U_0a/\nu$) |
| $Rm$ | Magnetic Reynolds number ($Rm=U_0a/\eta$) |
| $Pm$ | Magnetic Prandtl number ($Pm=Rm/Re$) |
| $\mathbf{u}^{k_x,k_y}$ | Flow with wavenumber $(k_x,k_y)$ |
| $\mathbf{b}^{k_x,k_y}$ | Magnetic field with wavenumber $(k_x,k_y)$ |
| $u_x^{0,k_y}$ | Zonal jets with wavenumber ($k_x=0$, $k_y$) |
| $\boldsymbol{\mathcal{E}}$ | Electromotive force (EMF) |
| $\alpha$ | Parker's alpha-effect |
| $\beta$ | Turbulent (beta-) diffusion |
| $\Upsilon$ | Upsilon dynamo source (a measure of turbulent cross-helicity $\tilde{\mathbf{u}}\cdot\tilde{\mathbf{b}}$) |
| $\Omega$ | A measure of omega-effect $\mathbf{B}\cdot\nabla\,\mathbf{U}$, which contributes zero in this work |

## References for Methods

**Acknowledgements**: We thank the anonymous reviewers who offered insightful comments and constructive suggestions that strengthened the results of the article. For fruitful discussions, it is our pleasure to thank J. R. Beattie, A. M. Beloborodov, A. Bhattacharjee, K. J. Burns, F. Ebrahimi, R. Habegger, D. Lecoanet, B. Miquel, E. R. Most, J. S. Oishi, and B. Ripperda. This work used Anvil at Purdue University through allocation TG-PHY130027 from the Advanced Cyberinfrastructure Coordination Ecosystem: Services & Support (ACCESS) program, which is supported by National Science Foundation grants #2138259, #2138286, #2138307, #2137603, and #2138296. Staff support from Anvil and additional computing resources from Bridges-2 are gratefully acknowledged.

**Funding**: This material is based upon work funded by the National Science Foundation (NSF) under award 2409206 and Department of Energy (Grant No. DE-SC0022257) through the DOE/NSF Partnership in Basic Plasma Science and Engineering. A.E.F. is supported by an NSF Astronomy and Astrophysics Postdoctoral Fellowship under award AST-2402142.

**Author contributions**: B.T. led the project, conceived the idea for the reported dynamo, developed analytic theory and schematic diagrams, modified numerical codes, performed simulations, acquired and analyzed data, and wrote the first draft of the article. All authors contributed significantly to the research. All authors helped develop research ideas, participated in the discussion of research methods, interpreted data, and edited and reviewed the article.

**Competing interests**: The authors declare no competing interests.

**Data and materials availability**: Any data needed to evaluate the conclusions of this article are present in the article or Methods or a Zenodo repository[56]. Simulation files, post processing scripts, additional codes and materials, data used to produce figures, and details of numerical code implementation are available on a Zenodo repository (https://doi.org/10.5281/zenodo.17162239). Upon request, B.T. will share additional details on numerical codes and analysis methods.